\documentclass{emulateapj}
\usepackage{eso-pic}
\usepackage{graphicx}

\newcommand{\cmpss}{cm~s$^{-2}$}
\newcommand{\mps}{m~s$^{-1}$}
\newcommand{\mpstwo}{m~s$^{-2}$}
\newcommand{\kps}{km~s$^{-1}$}

\newcommand{\gpcmthree}{g~cm$^{-3}$}
\newcommand{\Msun}{${\rm M_\odot}$}

\newcommand{\Mjup}{${\rm M_J}$}
\newcommand{\xonb}{XO-5b}
\newcommand{\xon}{XO-5}
\newcommand{\Rjup}{${\rm R_J}$}
\newcommand{\vMs}{1.0}		
\newcommand{\eMs}{0.03}
\newcommand{\vRs}{1.11}		
\newcommand{\eRs}{0.09}
\newcommand{\sptype}{G8V}
\newcommand{\vrvK}{145} 
\newcommand{\ervK}{10}

\newcommand{\vDs}{270}		
\newcommand{\eDs}{25}

\newcommand{\vjd}{2454485.6664}	
\newcommand{\ejd}{0.0004}	
\newcommand{\vap}{0.0508}	
\newcommand{\eap}{0.0005}

\newcommand{\vperiod}{4.187732}	
\newcommand{\eperiod}{0.00002}
\newcommand{\vMp}{1.15}
\newcommand{\eMp}{0.08}	
\newcommand{\vRp}{1.15}
\newcommand{\eRp}{0.12}
\newcommand{\vincl}{86.8}
\newcommand{\eincl}{0.9}

\newcommand{\vAge}{8.5}
\newcommand{\eAge}{0.8}
\newcommand{\vFeH}{0.25}
\newcommand{\eFeH}{0.03}
\newcommand{\vgp}{22}
\newcommand{\egp}{5}

\slugcomment{Submitted to Astrophysical Journal}

\received{2008 May 15}
\begin{document}

\title{XO-5\lowercase{b}: A Transiting Jupiter-sized Planet With A Four Day Period}

\author{
Christopher~J.~Burke\altaffilmark{1}
P.~R.~McCullough\altaffilmark{1},
Jeff~A.~Valenti\altaffilmark{1},
Doug~Long\altaffilmark{1},
Christopher~M.~Johns-Krull\altaffilmark{2},
P.~Machalek\altaffilmark{1,3},
Kenneth~A.~Janes\altaffilmark{4},
B.~Taylor\altaffilmark{4},
Michael~L.~Fleenor\altaffilmark{5},
Cindy~N.~Foote\altaffilmark{6},
Bruce~L.~Gary\altaffilmark{7},
Enrique~Garc\'{i}a-Melendo\altaffilmark{8},
J.~Gregorio\altaffilmark{9},
T.~Vanmunster\altaffilmark{10}
}

\email{cjburke@stsci.edu}

\altaffiltext{1}{Space Telescope Science Institute, 3700 San Martin Dr., Baltimore, MD 21218}
\altaffiltext{2}{Dept. of Physics and Astronomy, Rice University, 6100 Main Street, MS-108, Houston, TX 77005}
\altaffiltext{3}{Johns Hopkins University, Dept. of Physics and Astronomy, Baltimore, MD 21218}
\altaffiltext{4}{Boston University, Astronomy Dept., 725 Commonwealth Ave.,Boston, MA 02215}
\altaffiltext{5}{Volunteer Observatory, Knoxville, TN}
\altaffiltext{6}{Vermillion Cliffs Observatory, Kanab, UT}
\altaffiltext{7}{Hereford Arizona Observatory, Hereford, AZ}
\altaffiltext{8}{Esteve Duran Observatory Foundation, Montseny 46, 08553 Seva, Spain}
\altaffiltext{9}{Atalaia, Portugal}
\altaffiltext{10}{CBA Belgium Observatory, Landen, Belgium}

\begin{abstract}
The star \xon\ (GSC 02959-00729, V=12.1, \sptype) hosts a
Jupiter-sized, $R_{\rm p}$=\vRp$\pm$\eRp\ \Rjup, transiting extrasolar
planet, \xonb, with an orbital period of \vperiod$\pm$\eperiod\ days.
The planet mass ($M_{\rm p}$=\vMp$\pm$\eMp\ \Mjup) and surface gravity
($g_{\rm p}$=22$\pm$5 \mpstwo) are significantly larger than expected
by empirical $M_{\rm p}$-P and $M_{\rm p}$-P-[Fe/H] relationships.
However, the deviation from the $M_{\rm p}$-P relationship for \xonb\
is not large enough to suggest a distinct type of planet as is
suggested for GJ 436b, HAT-P-2b, and XO-3b.  By coincidence \xon\
overlies the extreme H I plume that emanates from the interacting galaxy
pair NGC 2444/NGC 2445 (Arp 143).
\end{abstract}

\keywords{planetary systems -- stars: individual (GSC 02959-00729) -- galaxies: individual (NGC 2444, NGC 2445)}

\section{Introduction}\label{sec:intro}

We report the discovery of a Jupiter-sized planet, \xonb, that transits the
\sptype, V=12.1, star GSC 02959-00729 (\xon) with an orbital period, P$\sim$
4 days.  \xonb\ provides a valuable
addition to the empirical population trends amongst transiting extrasolar
planets \citep{TOR08}, especially at P$>$4 day periods where only four
other well characterized transiting planets are known
\citep[OGLE-TR-111b, HAT-P-1b, HAT-P-2b, \& HD 17156b, ][
respectively]{PON04,BAK07,BAK07B,BAR07}.  With a sample of the first
six known transiting planets, \citet{MAZ05} pointed out the
possibility of a linearly decreasing trend of planet mass as a
function of increasing orbital period.  \citet{TOR08} and
\citet{SOU08} have recently revisited this and other relationships
amongst the now larger sample of transiting planets analyzed in a
homogeneous fashion.  As shown in Section~\ref{sec:disc}, \xonb\ is
one of the few planets that is inconsistent with general trends based
on the currently known sample of transiting Jupiter-sized planets.

In Section~\ref{sec:obs} we describe the photometric and spectroscopic
observations that we used to discover and characterize the \xon,
\xonb\ system.  Section~\ref{sec:sme} describes our spectroscopic
determination of the atmospheric properties of \xon, which we combine
with theoretical isochrones to constrain the stellar mass and radius.
The mass estimate of \xon\ is employed along with a high precision
transit light curve to measure the radius of \xonb\ in
Section~\ref{sec:lcmcmc}.  Radial velocity measurements determine the
mass of \xonb\ as outlined in Section~\ref{sec:rv}.  In
Section~\ref{sec:ttv}, the database of transit light curves enables
refining the transit ephemeris for \xonb.  We conclude in
Section~\ref{sec:disc} how the discovery of \xonb\ adds to the
understanding of the population of transiting Jupiter-sized planets.

\section{Observations}\label{sec:obs}

\subsection{XO Survey Photometry}\label{sec:xophot}

\citet{MCC05} provides details of the XO survey; additional details
of the candidate transiting planet selection process can be found in
\citet{MCC07}.  \xonb\ is the fifth Hot Jupiter transiting planet
announced by the XO survey \citep{MCC06,BUR07,JOH08,MCC08}, and was
identified as a candidate from the 62$\degr\times 7\degr$ field of
view centered on 8$^{\rm h}$ RA.  This field was observed at a 10 min
cadence with an overall rms noise, $\sigma=0.011$ mag, for this V=12.1
star during two seasons: Nov. 2003-Mar. 2004 and Nov. 2004-Mar. 2005.
\xon\ is well isolated; \xon\ contributes 90\% of the flux in the
75$\arcsec$ radius aperture used for XO photometry.

\begin{figure}
\plotone{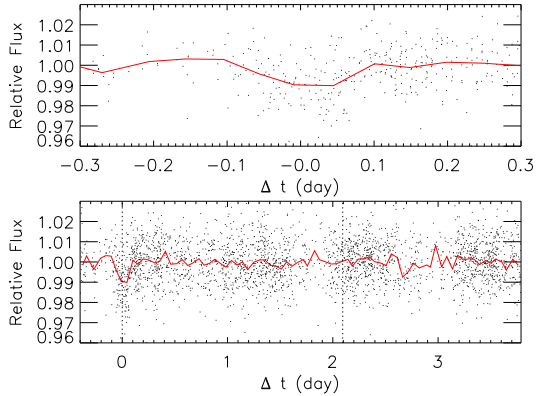}
\caption{Discovery phased light curve from the XO Project data.    ({\it Top}) The phased light curve around the transit event.  ({\it Bottom}) The phased light curve over the full orbital period.  The individual measurements ({\it points}) are shown binned ({\it solid line}) to reduce noise.  The transit occurs at $\Delta t=0.0$ day and any secondary eclipse for a circular orbit would occur at $\Delta t=2.1$ day ({\it dotted line}).  No secondary eclipse is evident above the noise.\label{fig:xophased}}
\end{figure}

Figure~\ref{fig:xophased} shows the XO light curve phased to the
period with the most significant transit event identified by the
Box-fitting Least Squares algorithm \citep{KOV02}.  In the phased
light curve, two full transit events and five partial events
contribute to the detection.  After subtracting a transit model
(determined from the high precision light curve in
Figure~\ref{fig:lowell}), the average residual RMS during transit
nights is $\sigma=0.008$ mag (30\% lower noise than the average for
the entire light curve).  Adding in quadrature the Signal to Noise
Ratio (SNR) for each transit yields an overall detection
SNR$_{Total}\sim 12$.  It is not routine to follow up all candidates
with this low of SNR$_{Total}$.  However, SNR$_{Total}$ is only one of
several criteria for selecting candidates
\citep{MCC07,BUR07}.  The efficiency of photometric followup provided
by the XO Extended Team is an additional resource of the XO
project to aid in following up these lower SNR$_{Total}$ candidates.

\subsection{XO Extended Team and Follow Up Photometry}\label{sec:etphot}

\begin{figure}
\plotone{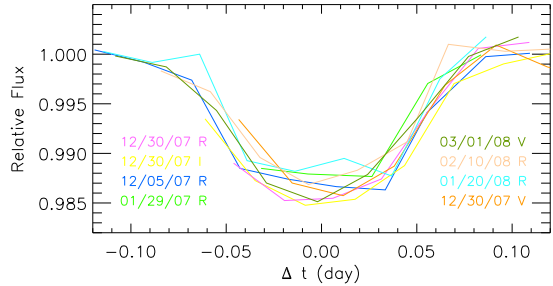}
\caption{Binned light curves from the Extended Team for \xon.  The text in color matches the corresponding light curve and indicates the date and passband of the observations.\label{fig:et}}
\end{figure}

The Extended Team (E.T.) provides photometric follow up of XO
candidates.  The E.T.\ (M.~F., C.~F., E.~G-M., J.~G., F.~M., G.~M.,
and T.~V.) is a collaboration of professional and amateur astronomers
\citep{MCC05,MCC06}.  The E.T. obtains light curves of XO candidates
at higher angular resolution and guides the photometric and
spectroscopic follow up necessary to classify a candidate as a bona
fide planetary companion.  Table~\ref{tab:lc} provides E.T. photometry
for \xon.  For the E.T. light curves, the median differential
magnitude out of transit provides the flux normalization and the
standard deviation out of transit provides the uncertainty in the
measurements.  Figure~\ref{fig:et} shows the E.T. light curves
obtained for \xonb.

\begin{figure}
\plotone{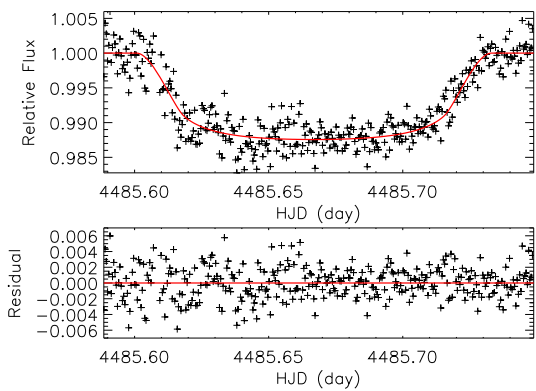}
\caption{{\it Top}: Light curve from the 1.8m Perkins Telescope at Lowell Observatory in the R-band ({\it points}), along with the best-fit transit model in a $\chi^{2}$ sense during the MCMC analysis ({\it solid line}). {\it Bottom}: Residual of data from the best-fit transit model.\label{fig:lowell}}
\end{figure}

On 2008 January 20 we observed a transit event of \xonb\
with the 1.8-m Perkins Telescope at Lowell Observatory using the PRISM
instrument in imaging mode \citep{JAN04}.  The observations are a
continuous series of R-band images covering a 10.8$\arcmin\times
8.8\arcmin$ subarray field of view with 10s exposure time and 30s
readout resulting in an overall $t_{cad}=$40 s cadence.  The detector
has 0.36$\arcsec$ pixels, and the FWHM seeing was 1.8$\arcsec$.  The
target was observed over the airmass range, $1.12\leq X \leq 1.6$,
with observations commencing at high airmass.  The evening was
nonphotometric; several percent transparency variations in raw
photometric magnitudes exist around the linear trend with airmass.  To
remove the transparency variations a differential light curve is
calculated employing four other comparable bright stars as comparison.
The resulting differential light curve of \xon\ is shown in
Figure~\ref{fig:lowell}.  The out of transit data are limited in
duration.  Thus, minimizing the scatter during the flat, in-transit
portion of the light curve, guided the choice of photometric aperture
size and comparison stars.  The resulting noise during the
in-transit portion is $\sigma_{exp}=0.002$ mag for each exposure.
This is $\sim 20\%$ higher noise than expected from Poisson,
scintillation \citep{DRA97}, sky, and read noise.  More relevant for
transit detection and characterization is the photometric noise rate,
$\sigma_{pnr}=\sigma_{exp}\sqrt{t_{cad}/60 s}=0.0016$ mag mn$^{-1}$,
which takes into account the dead time of detector readout.  The
calibration and photometry was performed in IRAF with a customization
in the IRAF phot task to report magnitudes to 10$^{-4}$ precision,
rather than the default of 10$^{-3}$ precision.

We obtained photometric B, V, ${\rm R_C}$, and ${\rm I_C}$ magnitudes
for \xon\ using a 0.35-m telescope (Table~\ref{tab:star}).
Photometric observations were obtained on three separate nights with
2-6 Landolt star fields \citep{LAN92} interspersed with the \xon\
field in order to define the color and airmass transformation from the
instrumental system to the standard system.  Each night 20-40 Landolt
stars defined the transformation with rms scatter of 0.04, 0.03, 0.02,
0.03 for the $BVRI$ passbands, respectively.  The resulting standard
photometry in Table~\ref{tab:star} is the weighted average amongst the
three nights.  Standard star photometry of \xon\ from {\it TASS}
\citep{DRO06} agrees with our photometry. Photometry from {\it 2MASS}
\citep{SKR06} is also given in Table~\ref{tab:star}.  \xon\ is
saturated in {\it SDSS} \citep{ADE08} images and too faint for the
{\it TYCHO-2} \citep{HOG00} database.

\subsection{Arp 143 - Potential Archival Measurements}\label{sec:arp143}
\begin{figure}
\plotone{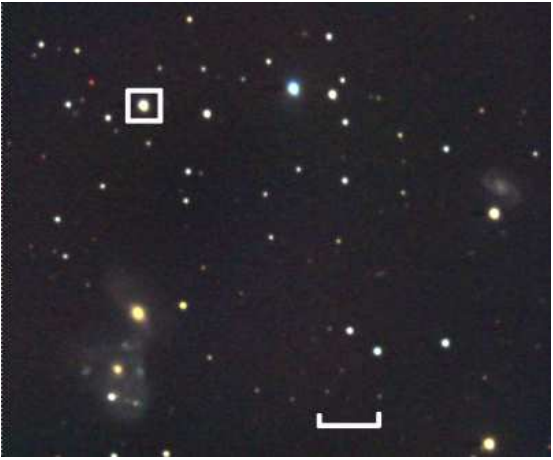}
\caption{False color image for finding \xon.  The box encloses \xon, and the scale bar indicates 1 arcmin.  The finder is $\sim$9 arcmin per side, and North is up and East is toward the left.  The interacting galaxy pair Arp 143 is visible in the lower left.\label{fig:finder}}
\end{figure}

\begin{figure}
\plotone{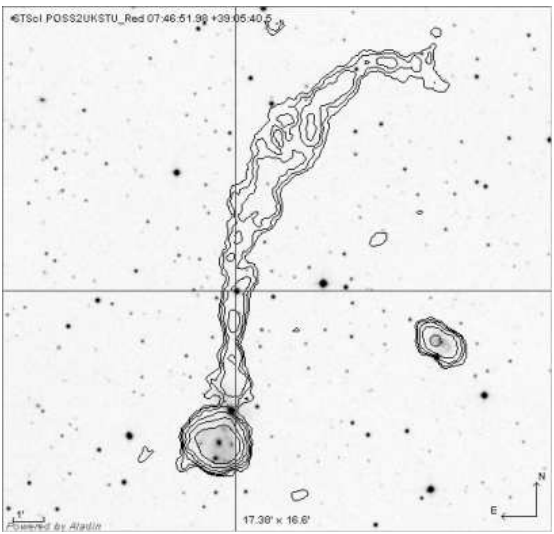}
\caption{Digital Sky Survey image (Red band) of \xon\ ({\it arrow}) showing the coincidental H I plume ({\it contour}) that emanates from the interacting galaxies Arp 143.  The H I emission contours are VLA observations from \citet{APP87}.\label{fig:HI}}
\end{figure}

By coincidence, \xon\ lies near the line of sight to the interacting
galaxy pair NGC 2444/5 cataloged as Arp 143 in the 'Material Emanating
from E Galaxies' group \citep{ARP66}.  Figure~\ref{fig:finder} shows a
false-color finder chart for \xon\ enclosed by the box along with the
Im peculiar galaxy NGC 2445 (patchy blue galaxy) and smoother S0
peculiar galaxy NGC 2444 three arcmin South of \xon.  The VLA has
detected one of the largest known (14.5$\arcmin$ in extent) H I
plumes, which emanates from Arp 143 that directly coincides with \xon\
as shown in Figure~\ref{fig:HI} \citep{APP87,HIG97,HIB01}.  With a
velocity shift of 4000 \kps, it is improbable that the H I emission
from Arp 143 can be attributed to XO-5.  Archival observations of Arp
143 may provide long term photometric monitoring or serendipitous
transit data for \xon, possibly constraining ephemeris variability.

\subsection{Spectroscopy}\label{sec:spec}
After confirmation of the XO transit light curve from E.T.\
observations (see \S\ref{sec:etphot}), we initiated queue schedule
observations of \xon\ with the High-Resolution Spectrograph (HRS), a
fiber fed cross-dispersed echelle spectrograph
\citep{TUL98}, on the 11-m Hobby-Eberly Telescope
(HET) located at McDonald Observatory in order to measure the mass of
the planet.  HRS observations with an iodine gas cell to yield precise
radial velocities commenced on 2007 December 7.  Table~\ref{tab:rv}
provides dates of the HRS observations along with the resulting radial
velocities.  The instrument setup provides R=63,000 resolution and
covered the wavelength range $4000<\lambda<7800$ $\AA$,
centered at $\lambda=5900$ $\AA$.  The two-dimensional echelle spectra
are extracted using procedures described in \citet{HIN00}.  Radial
velocities for \xon\ are determined in \S~\ref{sec:rv}.

To measure the stellar parameters of \xon, we also obtained HRS HET
spectra without the iodine absorption cell, using the 5936 \AA\ and
6948 \AA\ settings of the cross-disperser to cover the wavelength
intervals 5150 $< \lambda <5200$\AA\ and 6000 $< \lambda <6200$\AA\ at
R=63,000.  Three spectra at each configuration were co-added for
analysis.  The SNR is 35 and 50 for the co-added spectrum in the blue
and red configurations, respectively.  The \xon\ stellar parameters
are determined in \S~\ref{sec:sme}.

\section{Analysis}

\subsection{Stellar Properties}\label{sec:sme}

A transit light curve and radial velocity measurements do not
completely solve the system of stellar and planet properties; these
measurements determine the stellar density \citep{SEA03}.  Solving for the
stellar and planet properties independently requires an additional
constraint.  Following the procedure of \citep{BUR07} in analyzing the
transiting planet XO-2b, spectroscopic analysis combined with stellar
isochrones provides an estimate of the stellar mass, $M_{\star}$, and
its uncertainty.  This prior constraint on $M_{\star}$ enables
solving the system completely.  We used the Spectroscopy Made
Easy (SME) analysis package \citep{VAL96} with refinements from
\citet{VAL05} to determine the spectroscopic the properties of
\xon.

Table~\ref{tab:star} lists \xon\ stellar parameters from the SME
analysis of a single coadded spectrum.  The 1-$\sigma$ uncertainties
in the stellar parameters given in Table~\ref{tab:star} are based on
the typical rms scatter in parameters measured in independent,
multiple spectra for stars in the SPOCS catalog.  \xon\ has an
enhanced metal abundance, [Fe/H]=\vFeH$\pm$\eFeH, and the effective
temperature, $T_{eff}$ and surface gravity, log$g$ are consistent with
a \sptype\ spectral type according to Appendix B of \citet{GRA92}.
The empirical $T_{eff}$, intrinsic color calibrations of \citet{WOR06}
predict a color excess, $E(B-V)=0.03\pm 0.04$ and $E(V-K)=0.03\pm
0.03$ for \xon.  This difference is negligible within the
uncertainties, and a typical reddening law predicts
$E(V-K)_{pred}=2.74\times E(B-V)=0.08$, lower than the measured excess
$E(V-K)$.  The extinction map of \citet{SCH98} also measures a modest
$E(B-V)=0.05$ toward \xon.  During the analysis we assume that the
standard photometry is free of reddening.

\begin{figure}
\plotone{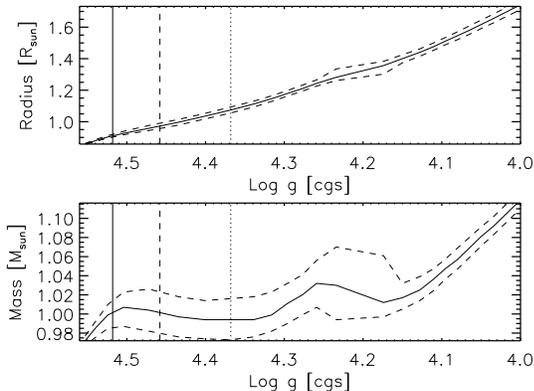}
\caption{Parametric curves for the Radius ({\it Top}) and Mass ({\it Bottom}) of \xon\ as a function of the surface gravity from the SME isochrone analysis, respectively (see \S~\ref{sec:sme}).  The SME spectroscopic determination of surface gravity for \xon\ ({\it solid}), the 1-$\sigma$ internal uncertainty ({\it dashed}), and the 1-$\sigma$ systematic uncertainty ({\it dotted}) are shown with vertical lines.\label{fig:sme}}
\end{figure}

Using the primary observables from the SME analysis ($T_{eff}$,
abundances, and log$g$) and the apparent V-band magnitude
(\S~\ref{sec:etphot}), we determined $M_{\star}$ and $R_{\star}$ from
Y$^{2}$ isochrones \citep{YI01}, using the procedure of \citet{VAL05}.
$M_{\star}$ and $R_{\star}$ depend on the unknown distance to \xon,
thus the probability density function for $M_{\star}$, $R_{\star}$,
and age are calculated for a sequence of trial distances in steps of
10 pc.  Estimates of $M_{\star}$ and $R_{\star}$ from this isochrone
analysis provide an additional estimate of log$g_{iso}$ as a function
of distance to \xon.  Figure~\ref{fig:sme} shows $R_{\star}$ and
$M_{\star}$ as a function of log$g_{iso}$.  A lower mass, lower
luminosity, younger age, and smaller distance star for \xon\
corresponds to higher surface gravity (left side of
Figure~\ref{fig:sme}); whereas, a higher mass, higher luminosity,
older age, and larger distance for \xon\ corresponds to lower surface
gravity (right side of Figure~\ref{fig:sme}).

A primary goal of the SME analysis is to provide an estimate of
$M_{\star}$ and its uncertainty in order to provide a unique solution
when fitting the light curve and radial velocity data.  The spectrum
alone yields an estimate of the stellar gravity, log$g_{sme}$ that is
independent of log$g_{iso}$.  The condition log$g_{sme}$=log$g_{iso}$
and the uncertainty in log$g_{sme}$ directly yields $M_{\star}$ from
the data shown in Figure~\ref{fig:sme}.  The solid vertical line in
Figure~\ref{fig:sme} shows log$g_{sme}$ which corresponds to
$M_{\star}=1.0 M_{\odot}$.  The long dashed vertical line in
Figure~\ref{fig:sme} represents the uncertainty of log$g_{sme}$,
$\sigma_{lgg}=0.06$ dex.  This uncertainty is an internal precision of
objects in the SPOCS catalog analyzed in a uniform manner.  For the
purpose of estimating $M_{\star}$, a more appropriate uncertainty in
log$g$ is determined through comparison of log$g_{sme}$ from the SPOCS
catalog to independent log$g$ measurements from the literature.
\citet{VAL05} find rms scatter $\sigma_{lgg}=0.15$ dex by comparing
the SPOCS catalog to external catalogs.  The dotted vertical line in
Figure~\ref{fig:sme} shows this larger uncertainty in log$g$.
$M_{\star}$ varies weakly with log$g$, and we adopt an uncertainty of,
$\sigma_{M}=0.03 M_{\odot}$ for $M_{\star}$, but also explore the
impact a larger, $\sigma_{M}=0.07 M_{\odot}$, has on the system
parameters.

\subsection{Markov Chain Monte Carlo Light Curve Analysis}\label{sec:lcmcmc}

\citet{FORD05}, \citet{GRE05}, and references therein provide a
thorough discussion of the theory behind Markov Chain Monte Carlo
(MCMC) Bayesian analysis along with a practical MCMC implementation
for radial velocity planet detection, and \citet{HOL06} introduced
MCMC techniques to the analysis of a transiting planet light curve.
To analyze the \xonb\ light curve, we follow the MCMC implementation
of \citet{BUR07} where they analyzed the light curve of the transiting
planet XO-2b.

To calculate the Bayesian posterior probability for the system
parameters, the likelihood function is given by $e^{-0.5\chi^{2}}$,
where we have assumed the errors are normally distributed, and the data
have uniform weights.  $\chi^{2}$ is the squared difference between
observations and the analytic transit model of \citet{MAN02}.  The
model assumes negligible eccentricity.  The observations are the
R-band data from the 1.8m Perkins Telescope shown in Figure~\ref{fig:lowell}.
To provide the additional constraint necessary to uniquely determine
the planet properties, we adopt an informative prior on $M_{\star}$
that is a Gaussian with center $M_{\star}=1.0 M_{\odot}$ and standard
deviation $\sigma=0.03 M_{\odot}$ (see \S~\ref{sec:sme}).  We adopt
uninformative priors for $R_{\star}$, $R_{p}$, and $i$ that are
uniform.  To improve the efficiency of the MCMC calculation, the computation is done using the set of parameters $R_{\star}$,
$\rho=R_{p}/R_{\star}$, and the total transit duration from 1$^{\rm
st}$ to 4$^{\rm th}$ contact, $\tau$ instead of $R_{\star}$, $R_{p}$,
and $i$ \citep{BUR07}.  Appendix A of \citet{BUR07} provides the form
of the priors necessary to maintain the uniform priors in $R_{\star}$,
$R_{p}$, and $i$ when the calculation is done using $R_{\star}$,
$\rho$, and $\tau$.  The limb darkening coefficients,$u_{1}$ and
$u_{2}$, of the transit model are allowed to vary.  Limb darkening is
described by a quadratic law, $I=1-u_{1}(1-\mu)-u_{2}(1-\mu)^{2}$,
where $I$ is the specific intensity normalized to unity at the center
of the stellar disk and $\mu$ is the cosine of the angle between the
line of sight and the surface normal.  In practice, we follow
\citet{HOL06} by adopting $a_{1}=u_{1}+2u_{2}$ and
$a_{2}=2u_{1}-u_{2}$ as the parameters used in the calculation.  The
limb darkening coefficients are constrained by requiring the highest
surface brightness to be located at the disk center ($u_{1}\geq 0.0$),
by requiring the specific intensity to remain above zero
($u_{1}+u_{2}\leq 1.0$), and by not allowing limb-brightened profiles
($u_{1}+2u_{2}\geq 0.0$).  The final free parameter is the mid-transit
time offset, $t_{o}$, from the initial ephemeris with a period
obtained from analyzing the XO observations and E.T. observations (see
\S~\ref{sec:ttv}.
  
The Markov Chain analysis employs the Metropolis-Hastings algorithm
with a normal proposal distribution and a multiple block sampling
technique where each step in the chain consists of a number of intra
steps updating each individual parameter in turn.  Several short,
trial chains iteratively yield scale factors of the normal proposal
distribution for each parameter with a 25\% to 40\% acceptance rate
for the trial samples.  In addition to the choice of $R_{\star}$,
$\rho$, and $\tau$ as the variables for the calculation, a linear
transformation between parameters yields an eigenbasis set of
parameters with a multi-normal non-covariant relationship that further
reduces the mutual degeneracy amongst the parameters \citep{BUR07}.  These
transformations yield an efficient MCMC calculation with an
autocorrelation length $N_{cor}=5$ for the chain.  The estimate of the
posterior probability comes from 7 independent chains of length
$N_{chn}=70000$ with varying initial conditions.  This results in an
effective length $N_{eff}=N_{chn}/N_{cor}=14000$.

\begin{figure}
\plotone{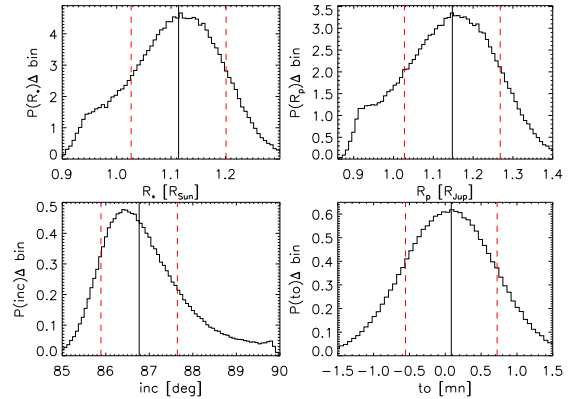}
\caption{Marginalized posterior probability for the \xon\ and \xonb\ parameters from MCMC samples.  We adopt the median of the posterior probability ({\it solid}) vertical line as the point estimate of the parameter.  The {\it dashed} lines indicate the 68\% credible interval.\label{fig:params}}
\end{figure}

\begin{figure}
\plotone{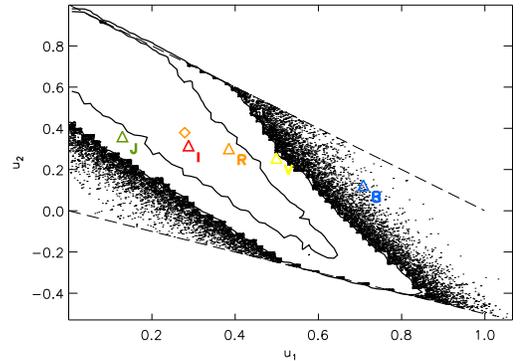}
\caption{ Joint posterior probability for the stellar limb darkening coefficients, $u_{1}$ and
$u_{2}$.  The solid contours are isoprobability contours containing
68\% and 90\% of the MCMC samples.  The remaining 10\% of the samples
lying outside the region of highest probability are also shown ({\it
points}).  The theoretically calculated limb darkening coefficients
are shown for various passbands ({\it triangle}) along with the best
estimate limb darkening coefficients from the MCMC analysis of the
R-band transit light curve ({\it diamond}).  The R-band light curve
coefficients are consistent with the theoretically calculated R-band
coefficients.  The prior limits for the limb darkening coefficients
are indicated with {\it dashed lines}.\label{fig:limbcontour}}
\end{figure}

Figure~\ref{fig:params} shows the resulting posterior probability
distribution for select system parameters after marginalization over
the other parameters.  The posterior probability is simply a
normalized histogram of the MCMC sample values.  We adopt the median
as the best single point estimate of the posterior probability. The
1-$\sigma$ credible interval for a parameter is given by the
symmetrical interval around the median that contains 68\% of the
samples.  Figure~\ref{fig:limbcontour} shows the joint posterior
probability for the stellar limb darkening coefficients, $u_{1}$ and
$u_{2}$.  The constraints on the limb darkening coefficients are not
very constraining.  Figure~\ref{fig:limbcontour} shows theoretically
calculated limb darkening coefficients in various photometric
passbands from \citet{CLA00} for a star with the physical properties
of \xon.  The best estimates for limb darkening coefficients (diamond
symbol) from the R-band transit light curve is consistent with the
theoretical R-band coefficient (triangle symbol).

Using the SME isochrone analysis, the MCMC samples for $R_{\star}$
translate into distance to and age estimates of \xon.  Similar to the
procedure that defines the prior on $M_{\star}$, The SME isochrone
analysis provides a parametric relationship between $R_{\star}$ versus
trial distance and age (i.e., for a given stellar radius estimate, the
SME isochrone analysis has a best distance and age estimate for the
system).

The theoretical planet models of \citet{FOR07} characterize
the amount of heavy elements (elements heavier than H \& He) for a
given orbital semi-major axis, $R_{\rm p}$, and $M_{\rm p}$.  The best
estimate of $R_{\rm p}$ for \xonb\ is consistent with zero heavy
element content, and the 1-$\sigma$ upper limit is 5 $M_{\earth}$ of
heavy elements.  Table~\ref{tab:planet} summarizes the properties of
\xonb.

The uncertainties given in Table~\ref{tab:planet} for \xonb\ and
Table~\ref{tab:star} for \xon\ represent the internal precision of the
experiment.  These uncertainties represent the expected scatter of
values obtained if the experiment was repeated with similar quality
data and identical procedures.  Other systematic sources of error are
most likely comparable or larger than the internal precision.  The
sources of systematic error only enter into the prior for $M_{\star}$.
In light of these potential sources of systematic uncertainty, the
analysis is repeated with an increased standard deviation of the
Gaussian prior on $M_{\star}$ to $\sigma=0.07$ \Msun.  This larger
uncertainty on $M_{\star}$ did not increase the uncertainties in the
other parameters beyond the most significant digit as given in
Table~\ref{tab:star} and Table~\ref{tab:planet}.  This indicates the
photometric noise rather than uncertainty in $M_{\star}$ dominates the
uncertainty in the system parameters.

After fixing the prior on $M_{\star}$, the light curve analysis yields
an additional estimate of log$g_{mcmc}$ as given in
Table~\ref{tab:star}.  log$g_{mcmc}$ does not differ significantly
from the SME based log$g_{sme}$.  We redetermined the stellar
properties by fixing log$g_{sme}$=log$g_{mcmc}$ when analyzing the
spectra.  This reduced $M_{\star}=0.02$ \Msun.  This variation is
within our original uncertainty of $M_{\star}$.  Thus, we did not iterate the MCMC analysis with the SME analysis \citep{SOZ07}.

\subsection{Radial Velocity Measurements}\label{sec:rv}

We determined the mass of \xonb\ using the radial velocity techniques
described by \citet{MCC06}, \citet{BUR07}, and \citet{JOH08}.  We
summarize the procedure here, where we derive radial velocity
information from the HET spectra (see
\S~\ref{sec:spec}).  An iodine gas cell imprints iodine absorption lines
on the spectrum of \xon\, and this is compared to a model spectrum to
obtain radial velocity shifts with respect to the topocentric frame.
We construct model spectra by multiplying a very high-resolution FTS
spectrum of the Sun (scaled to better match the observed line depths
of \xon) times a very high-resolution FTS spectrum of an iodine gas
cell \citep{COC00} and then convolving the result with a Voigt profile
to approximate the line-spread function of the instrument.  A radial
velocity shift is determined independently for each $\sim$15 \AA\ row
of the 2-D echelle spectrum of \xon\ over the wavelength range with
strong iodine absorption, 5210 $< \lambda <$ 5700 \AA.  The individual
stellar radial velocity estimates from each of the 15 \AA\ sections
(after transformation to the barycentric frame) determines the stellar
radial velocity measurement at each epoch and its associated
1-$\sigma$ uncertainty (see Table~\ref{tab:rv}).

\begin{figure}
\plotone{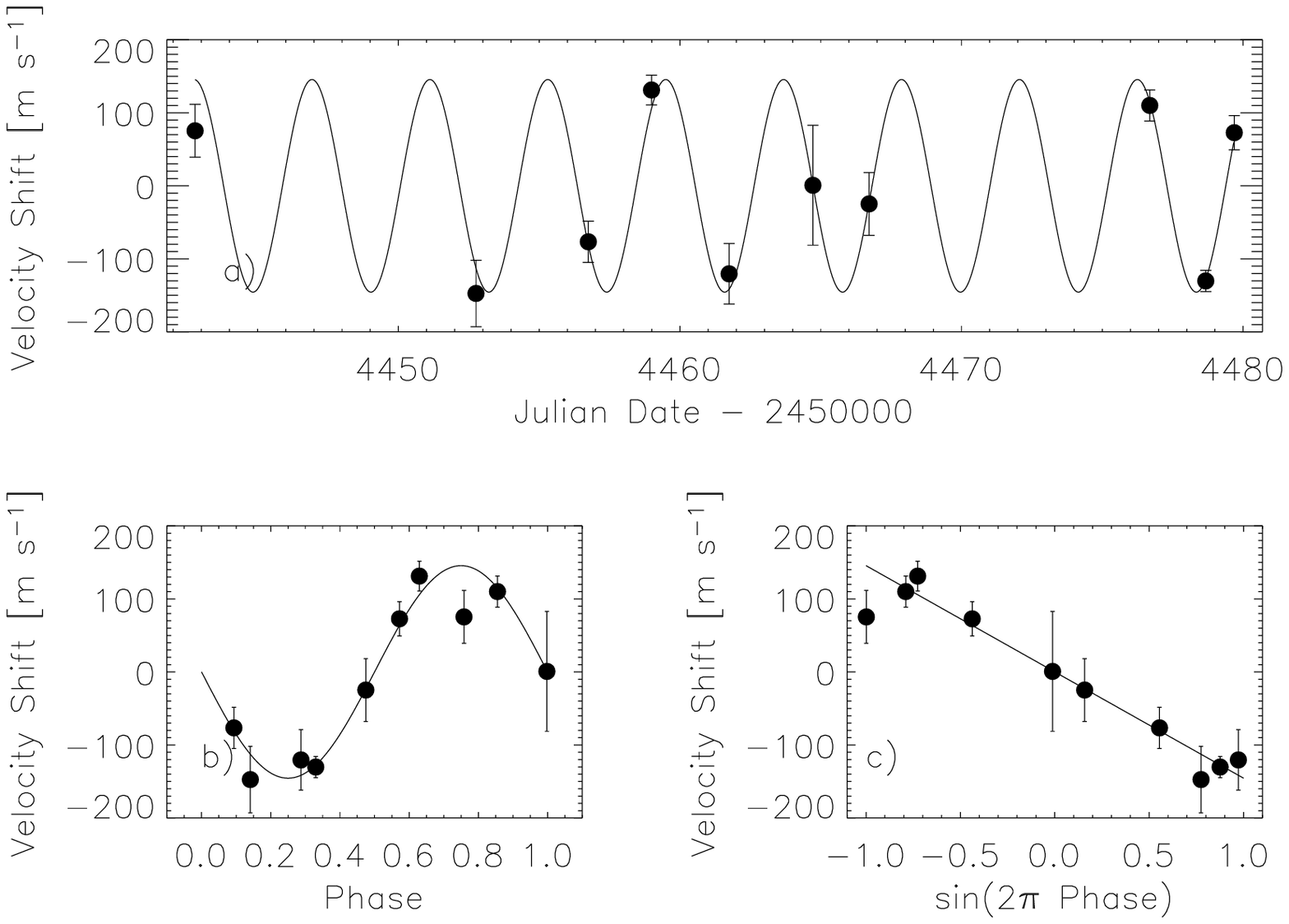}
\caption{a) The radial velocity of \xon\ oscillates sinusoidally with a
semi-amplitude K = \vrvK$\pm$\ervK\ \mps, implying \xonb's mass is
\vMp$\pm$\eMp\ \Mjup. b) The
period and phase of the radial velocities were fixed at values
determined by the transits. The mean stellar radial velocity with
respect to the solar system's barycenter has been subtracted.  In
order to determine K, we used the HET spectra calibrated with an
iodine absorption cell (filled circles).   c) In this representation
of the data, a circular orbit yields a straight line of slope $-$K.
\label{fig:rv}}
\end{figure}

Figure~\ref{fig:rv} shows the resulting radial velocity curve phased
with the \xonb\ ephemeris determined from the transits and assuming
zero eccentricity.  The typical uncertainty per epoch is
$\sigma_{RV}=20-40$ \mps, limited mainly by Poisson noise in the
spectrum.  The radial velocity semi-amplitude, K=\vrvK$\pm$\ervK\
\mps.  This amplitude results in $M_{p}=$\vMp$\pm$\eMp\ \Mjup\ for
\xonb, assuming $M_{\star}$=\vMs\ \Msun\ for \xon\ and a circular orbit for \xonb.

\subsection{Ephemeris and Transit Timing Variations}\label{sec:ttv}

The XO, E.T., and Perkins Telescope transit light curves enable
refining the ephemeris for \xonb.  The high precision light curve from
the Perkins Telescope (see \S~\ref{sec:etphot}), provides the estimate
for the mid-transit zero point of the ephemeris as well as the transit
model employed to derive mid-transit timing for the other lower
precision light curves from the E.T. and XO survey.  The MCMC samples
from the Perkins Telescope light curve analysis (lower right panel of
Figure~\ref{fig:params}) provide the best estimate and error for the
ephemeris zero point, $HJD_{o}=$\vjd$\pm$\ejd\ ($\sigma=35$ s).

To refine the orbital period of \xonb, the remaining XO and E.T.\
light curves (E.T.\ light curves are shown in Figure~\ref{fig:et})
provide mid-transit timing estimates.  The best fitting transit model
in a $\chi^{2}$ sense to the high precision Perkins Telescope light curve
(Figure~\ref{fig:lowell}) is fit to the other light curves with only
mid-transit time and out of transit flux zero point as the only free
parameters.  The expected variations in the transit model as a
function of photometric bandpass is negligible compared to the
photometric noise in the E.T.\ and XO light curves, so the limb
darkening coefficients are fixed to the values from analyzing
the R-band Perkins light curve.  Table~\ref{tab:midpoints} provides
mid-transit timings along with their uncertainty.  The XO observations
have 10 min cadence and $\sim$1\% photometric noise making transit
timing uncertain to 28 min.  This uncertainty comes from comparing
the a transit model fit a by-eye estimate
of the mid-transit timing.  Within the uncertainties there is not a
significant systematic difference between mid-transit timings
estimated by-eye and from the model fit.  The much higher quality
E.T.\ light curves have an uncertainty of 4.3 min in mid-transit
timing.  This uncertainty is estimated from comparing the groups of
mid-transit timing observations for the same event but different
observers.

With these uncertainties, an ephemeris model with period as the only
free parameter results in $\chi^{2}=12.6$ with $\nu=14$ degrees of
freedom indicating the timings are consistent with a fixed period,
$P=$\vperiod$\pm$\eperiod\ day.  The ephemeris for \xonb\ accumulates
a 5 min uncertainty by 2010.  Given the large uncertainty in transit
mid-point timing from the XO survey data, the period was calculated
using E.T.\ data alone.  The earliest, and only, E.T.\ light curve
from the 2006-2007 observing season for \xonb\ provides the strongest
constraint on the orbital period.  The remaining E.T.\ light curves
are from the 2007-2008 observing season for \xonb.  The period derived
from E.T.\ data alone is within 1-$\sigma$ of the determination that
includes the XO survey data.

\section{Discussion \& Conclusion}\label{sec:disc}
The transit candidate of \citet{MAN05} illustrates the non-negligible
potential for triple stars to have transit light curves and radial
velocity variations consistent with a planet.  However, in the case of
\xonb, attempts to explain the light curve and spectroscopy with a
physical stellar triple fail.  We employ the Y$^{2}$ isochrone
appropriate for the the physical properties of \xon\ supplemented with
the low-mass stellar isochrone between 0.072$<M_{\star}<$0.5 \Msun\
from \citet{CHAB00}, stellar limb darkening coefficients from
\citet{CLA00}, the light curve synthesis routine of \citet{WIL93}, and
assume any potential stellar binary has zero orbital eccentricity to
model a stellar triple system.  The constraints on the transit
duration and transit depth from the light curve require
$M_{\star}>0.90$ \Msun\ for the primary of a stellar binary blended
with the light of \xon.  The required primary of the stellar binary
has $>$45\% the flux of \xon\ and has a radial velocity
semi-amplitude, $K> 18$ \kps.  Such a binary would be readily apparent
in the spectrum of \xon\ given the narrow spectral features for \xon,
$v\sin{i}<3.3$ \kps.  We cannot completely rule out the possibility of
a line-of-sight faint background binary blended with the light of
\xon\ as an explanation for the observations.  However, the sinusoidal
shape of the radial velocity variations necessitate the line-of-sight
binary to have a systemic velocity similar to \xon\, otherwise the
radial velocity curve develops asymmetries that are not observed
\citep{TOR05}.

\citet{TOR08} and \citet{SOU08} recently derived transiting planet
properties in a uniform fashion in order to further test various
trends amongst the transiting planets that have previously been
examined with smaller and less homogeneous samples of transiting
planets \citep{MAZ05}.  The larger sample of transiting planets in
\citet{TOR08} still shows a general trend for decreasing $M_{\rm p}$
with increasing orbital period, but there is significant scatter and
some very significant outliers (GJ 436b, HAT-P-2b, and XO-3b).
Furthermore, \citet{TOR08} find a $M_{\rm p}$-P-[Fe/H] relation that
reduces the scatter in the $M_{\rm p}$-P relation.  At fixed period,
the more metal poor stars host higher mass planets.  \xonb\ has
$\Delta M_{\rm p}=0.6\pm$\eMp\
\Mjup\ higher mass than then $M_{\rm p}$-P trend defined in
\citet{TOR08}.  The metallicity correction to the $M_{\rm p}$-P trend found by \citet{TOR08}, $\Delta M_{\rm p}=-0.14$ \Mjup\ in the case of \xonb, is opposite sign from what is measured.  As commented by \citet{TOR08}, the $M_{\rm p}$-P-[Fe/H] relation
does have a scatter larger than observational uncertainties.  Thus,
additional physical conditions impact the $M_{\rm p}$-P relation, the
full variety of Jupiter-class extrasolar planets has not been fully
explored, or the trend is more relevant to shorter orbital periods
than \xonb.  For instance, \citet{MAZ05} propose the $M_{\rm p}$-P
relation is setup by the thermal evaporation of planets too small to
survive the stellar host XUV flux.

Related to $M_{\rm p}$, \citet{SOU07} and \citet{SOU08} show the
planet surface gravity, $g_{\rm p}$ is also correlated with P.  The
scatter of the $g_{\rm p}$-P relation is also larger than the
observational uncertainties.  \xonb\ has a higher $g_{\rm p}$ than the
trend based on previously known extrasolar planets.  It is not clear
which parameter $M_{\rm p}$ or $g_{\rm p}$ is more fundamentally
correlated, if at all, with P.  \citet{SOU07} suggests the $g_{\rm p}$-P
relation is more fundamental than the $M_{\rm p}$-P relation given the direct
role $g_{\rm p}$ has on the evaporation of highly irradiated gas giant
planets.  \citet{TOR08} note that the scatter around their $g_{\rm
p}$-P trend does not correlate with metallicity.

The previously known sample of transiting planets was shown by
\citet{HAN07} to separate into two classes by the Safronov
number of the planet.  The Safronov number,
$\theta=1/2(V_{esc}/V_{orb})^2$, where $V_{esc}$ is the escape
velocity from surface of the planet and $V_{orb}$ is the planet
orbital velocity, is a measure of the ability of a planet to
gravitationally scatter other bodies.  \citet{HAN07} and refined by
\citet{TOR08} find transiting planets fall into categories defined as
Class I, $\theta\sim0.07\pm0.01$ and Class II, $\theta\sim0.04\pm0.01$.
\xonb, $\theta=0.1\pm0.01$ does not appear to belong to either class,
yet it is not as discrepant as the other outlying transiting planets,
GJ 436b ($\theta=0.025$) and HAT-P-2b ($\theta=0.94$) \citep{TOR08}.
There is some overlap, but the planets with higher metallicity stellar
hosts tend to be the Class II objects with lower $\theta$, whereas
Class I objects have increasing $\theta$ toward lower metallicities.
\xonb\ does not follow this trend, but has high $\theta$ for higher
metallicity.  

A single object such as \xonb\ does not invalidate the trends amongst
the Jupiter-class of transiting planets, but illustrates the
importance of expanding the number of such objects to fully explore
the diversity of this population.  Recently, the SuperWASP project
announced 10 new transiting
planets\footnote{http://www.superwasp.org}.  We look forward to the
detailed analysis of these new transiting planets to amplify or
diminish the strength of the current trends amongst the close-in
transiting planets.  The end goal of a large population of transiting
planets is to investigate any trends that may directly constrain
processes that affect planet formation, migration, and planet
survival.

\acknowledgments

The University of Hawaii staff have made the operation on Maui
possible; we thank especially Jake Kamibayashi, Bill Giebink, Les
Hieda, Jeff Kuhn, Haosheng Lin, Mike Maberry, Daniel O'Gara, Joey
Perreira, Kaila Rhoden, and the director of the IFA, Rolf-Peter
Kudritzki.  The Hobby-Eberly Telescope (HET) is a joint project of the
University of Texas at Austin, the Pennsylvania State University,
Stanford University, Ludwig-Maximilians-Universit\"{a}t M\"{u}nchen,
and Georg-August-Universit\"{a}t G\"{o}ttingen.  The HET is named in
honor of its principal benefactors, William P. Hobby and Robert
E. Eberly.  We thank the HET night-time and day-time support staff and
the Resident Astronomer telescope operator; we especially thank John
Caldwell, Frank Deglman, Heinz Edelmann, Stephen Odewahn, Vicki Riley,
Sergey Rostopchin, Matthew Shetrone, and Chevo Terrazas.

We thank Dave Healy, Lisa Prato, and Naved Mahmud for assistance
observing; Jim Heasley for assistance on the XO survey; Ron Bissinger,
Paul Howell, Franco Mallia, and Gianluca Masi for their contributions
in following up XO candidates; and John Hibbard and Phil Appleton for
kindly making the H I radio data of Arp 143 available.

This research has made use of the SIMBAD database, operated at CDS,
Strasbourg, France; data products from the Two Micron All Sky Survey
(2MASS), the Digitized Sky Survey (DSS), and The Amateur Sky Survey
(TASS); source code for transit light-curves (Mandel \& Agol 2002);
and community access to the HET.

XO is funded primarily by NASA Origins grant NNG06GG92G and the Director's
Discretionary Fund of the STScI.

\clearpage

\begin{deluxetable}{cccccc}
\tabletypesize{\small}
\tablewidth{0pt}
\tablecaption{{\rm XO, E.T., \& Perkins Light Curve Data}\tablenotemark{a}}
\startdata
\hline
\hline
Heliocentric Julian Date & Light Curve & Uncertainty & Filter & N\tablenotemark{b} & Observatory \\
                         &  [mag]      & (1-$\sigma$) [mag] & & & \\
\hline
2452961.13135 & -0.0033 & 0.0113  &   W &  1 &  XO \\
2452961.13135 & -0.0095 & 0.0113  &   W &  1 &  XO \\
2452961.14526 &  0.0286 & 0.0113  &   W &  1 &  XO \\
2452961.14526 &  0.0048 & 0.0113  &   W &  1 &  XO \\
2452964.11816 &  0.0187 & 0.0100  &   W &  1 &  XO \\
\hline
\enddata
\tablenotetext{a}{The complete version of this table is in the electronic edition of
this article.  The printed edition contains only a sample.}
\tablenotetext{b}{Average of N measurements}
\label{tab:lc}
\end{deluxetable}

\begin{deluxetable}{lcl}
\tabletypesize{\small}
\tablewidth{0pt}
\tablecaption{{\rm Stellar Properties}}
\startdata
\hline
\hline
Parameter & Value of \xon & Reference\\
\hline
GSC ID        & 02959-00729                    & a \\
RA (J2000.0)  & $ 7^h46^m51^s.98 $             & a \\
Dec (J2000.0) & +39\arcdeg05\arcmin40\arcsec.5 & a \\
Galactic Latitude b [deg]       & 26.94        & a \\
  ''     Longitude l [deg]      & 180.63       & a \\
V             & 12.13$\pm$0.03                 & b \\
(B-V)         & 0.84$\pm$0.05                  & b \\
(V-R$_{c}$)     & 0.47$\pm$0.05                & b \\
(V-I$_{c}$)     & 0.83$\pm$0.05                & b \\
V$_{\rm TASS}$  & 12.17$\pm$0.06               & c \\
(V-I$_{c}$)$_{\rm TASS}$ & 0.82$\pm$0.08       & c \\
J             & 10.77$\pm$0.02                 & d \\
(J-H)         & 0.33$\pm$0.03                  & d \\
(H-K)         & 0.10$\pm$0.03                  & d \\
Spectral Type & \sptype                        & b \\
Distance [pc] & \vDs$\pm$\eDs                 & b \\
$\mu_{\alpha}$ [mas\ yr$^{-1}$] & -32.3$\pm$2.7 & e \\
$\mu_{\delta}$ [mas\ yr$^{-1}$] & -24.4$\pm$5.6 & e \\ 
Stellar Mass [$M_{\odot}$] &  \vMs$\pm$\eMs     & b,f \\
Stellar Radius [$R_{\odot}$] & \vRs$\pm$\eRs    & b \\
$T_{eff}$ [K]	  & 5510$\pm$44                 & b,f \\
\ [Fe/H]	          & 0.25$\pm$0.03                & b,f \\
log$g$ [\cmpss]	  & 4.52$\pm$0.06                & b,f \\
log$g$ [\cmpss]	  & 4.34$\pm$0.07                & b,g \\
$v$~sin~$i$ [\kps] & 1.8$\pm$0.5	         & b,f \\
\ [Na/H]	          & 0.18$\pm$0.03                & b,f \\
Stellar $\rho$ [\gpcmthree] & 1.02$\pm$0.2         & b \\
Age [Gyr]                 & \vAge$\pm$\eAge      & b \\  
\enddata
\\
References:\\
a) SIMBAD\\
b) this work \\
c) TASS \citep{DRO06} \\
d) 2MASS \citep{SKR06} \\
e) UCAC2 \citep{ZAC04} \\
f) SME Spectroscopic Determination (\S~\ref{sec:sme}) \\
g) Transit Light Curve Determination (\S~\ref{sec:lcmcmc}) \\
\label{tab:star}
\end{deluxetable}

\begin{deluxetable}{ccc}
\tabletypesize{\small}
\tablewidth{0pt}
\tablecaption{{\rm \xon\ Radial Velocity Shifts}}
\startdata
\hline
\hline
Julian Date & Radial Velocity &  Uncertainty \\
  -245000   &  Shift [\mps] &  (1-$\sigma$) [\mps]  \\
\hline
       4442.7789  &     75.4  &   36 \\
       4452.7562  &   -147.4  &   46 \\
       4456.7451  &    -76.6  &   28 \\
       4458.9892  &    131.2  &   20 \\
       4461.7447  &   -120.5  &   41 \\
       4464.7206  &      0.7  &   82 \\
       4466.7167  &    -24.8  &   43 \\
       4476.6851  &    110.0  &   21 \\
       4478.6743  &   -130.2  &   14 \\
       4479.6874  &     72.8  &   24 \\
\enddata
\label{tab:rv}
\end{deluxetable}

\begin{deluxetable}{ccc}
\tabletypesize{\small}
\tablewidth{0pt}
\tablecaption{{\rm Mid-Transit Times}}
\startdata
\hline
\hline
Heliocentric Julian Date & Uncertainty         & Observatory ID\tablenotemark{b} \\
 -2450000                &  (1-$\sigma$) [day] &                  \\
\hline
2999.0114 &  0.02  & XO \\
3354.9557 &  0.02  & XO \\
3359.1669 &  0.02  & XO \\
3375.9428 &  0.02  & XO \\
3417.8079 &  0.02  & XO \\
3438.7377 &  0.02  & XO \\
3442.9195 &  0.02  & XO \\
4129.7088 &  0.003  & MF \\
4439.6022 &  0.003  & EGM \\
4464.7268 &  0.003  & CF \\
4464.7331 &  0.003  & CF \\
4485.6730 &  0.003  &  CF \\
4485.6685 &  0.003  &  JG \\
4506.6061 &  0.003  &  MF \\
4527.5427 &  0.003  &  JG \\
\enddata
\tablenotetext{a}{Observatory ID is author initials, except XO is XO survey data.}
\label{tab:midpoints}
\end{deluxetable}

\begin{deluxetable}{lc}
\tabletypesize{\small}
\tablewidth{0pt}
\tablecaption{{\rm The Planet \xonb}}
\startdata
\hline
\hline
Parameter & Value \\
\hline
$P $ 				& \vperiod$\pm$\eperiod\ d 		\\
$t_c $	 			& \vjd$\pm$\ejd\ (HJD)			\\
$K $ 				& \vrvK$\pm\ervK$\ \mps 	 	 \\
$a $ 				& \vap$\pm$\eap\ A.U. 			 \\
$i $  		                & \vincl$\pm$\eincl\ deg  		 \\
$M_{\rm p} $ 			& \vMp$\pm$\eMp\ \Mjup		 	\\
$R_{\rm p} $ 			& \vRp$\pm$\eRp\  \Rjup       \\
$g_{\rm p} $                    & \vgp$\pm$\egp\           \mpstwo        \\
$a/R_{\star}$                   & 9.81$\pm$0.8 \\
$R_{\rm p}/R_{\star}$           & 0.106$\pm$0.003 \\
$\tau$                          & 3.13$\pm$0.07 hr \\
Impact parameter, b             & 0.55$\pm$0.09 \\
log$g_{\rm p}$                  & 3.35$\pm$0.09 cgs  \\
$\rho_{\rm p}$                  & 1.02$\pm$0.3 \gpcmthree \\
$T_{eq}=T_{eff}(R_{\star}/2a)^{1/2}$ & 1244$\pm$48 K \\
Safronov                        & 0.10$\pm$0.01 \\
\enddata
\label{tab:planet}
\end{deluxetable}

\end{document}